\documentstyle[12pt]{article}

\begin{document}

\title{Rotating Boson and Boson-Fermion Stars}
\author{Claudio M. G. de Sousa
\footnote{Present address: Universidade Catolica de Brasilia, Diretoria de Fisica, QS 07, Lt 01, EPCT - Aguas Claras, Brasilia, DF 72022-900, Brazil. ( claudio@unb.br )}\\
Universidade Federal do Tocantins, PROPESQ,\\
C.P. 111, Palmas, TO 77001--970, Brazil}

\maketitle

\abstract{Some recent results on rotating self-gravitating configurations
composed with Bosons and Fermions are reported. Given a star composed of
both Bosons and Fermions without interaction, it is shown that it is
possible to obtain stable slowly rotating configurations by using the same
perturbative relativistic method that usually describes neutron stars.}

\section{Introduction}

Fermion stars is a general name given to the well known neutron stars. Since
Oppenheimer and Volkoff \cite{Oppenheimer39}, these compact objects 
have received large
attention and many of their properties are already determined, and
astronomers can use the theoretical information to seek for those objects.
From Chandrasekhar \cite{Chandrasekhar39} one can also learn that Neutron stars can be
studied using the relativistic approach as a degenerate Fermi gas under
itself gravitational field, with an equation of state to determine its
pressure and density of energy.

In contrast to fermion star there is the so-called Boson Star, made up with
self-gravitating bosons at zero temperature. In this case, one can also use
relativistic approach, but the star structure can be studied directly from
the stress-energy tensor since it is possible to write a Lagrangian for
bosons in place of using an equation of state.

In this report I will show some recent results concerning the rotation of
boson star and of boson-fermion stars.

\section{Non-rotating Boson-Fermion Stars}

\subsection{Building Boson Stars}

Boson Stars have been originally proposed by
Ruffini \& Bonazzola \cite{Ruffini69}.

Metric of spherically symmetric configuration: 
\begin{equation}
ds^{2}=-Bd\tau ^{2}+Adr^{2}+r^{2}d\theta ^{2}+r^{2}\sin ^{2}\theta d\varphi 
\label{e1}
\end{equation}

The Lagrangian is: 
\begin{equation}
L=\frac{R}{16\pi G}-\partial _{\mu }\Phi ^{*}\partial ^{\mu }\Phi -m^{2}\Phi
^{*}\Phi   \label{e2}
\end{equation}

Where we use the ansatz: 
\begin{equation}
\Phi (r,\tau )=\phi (r)e^{-i\omega \tau }  \label{e3}
\end{equation}

For simplicity we are not going to consider scalar field self-interaction
\cite{Sousa98} and, the Klein-Gordon equation reads has no source term.

The energy-momentum tensor for bosons is: 
\[
T_{\mu \nu }^{B}=\partial _{\nu }\Phi ^{*}\partial _{\mu }\Phi +\partial
_{\mu }\Phi ^{*}\partial _{\nu }\Phi -g_{\mu \nu }\left( \partial _{\lambda
}\Phi ^{*}\partial ^{\lambda }\Phi +m^{2}\Phi ^{*}\Phi \right) 
\]

And conserved quantities:

\[
J^{\mu}=i g^{\mu\nu}(\Phi^{*} \partial_{\nu}\Phi - \Phi
\partial_{\nu}\Phi^{*} ) 
\]

\[
N_{B}=\int d^3 r \sqrt{-g} J^0 
\]

The set of differential equation reads:

\begin{equation}
A^{\prime }=xA^{2}\left[ \left( \frac{w^{2}}{B}+1\right) \sigma ^{2}+\frac{%
\sigma ^{\prime \,2}}{A}\right] -\frac{A}{x}(A-1)  \label{e4}
\end{equation}

\begin{equation}
B^{\prime }=xAB\left[ \left( \frac{w^{2}}{B}-1\right) \sigma ^{2}+\frac{%
\sigma ^{\prime \,2}}{A}\right] +\frac{B}{x}(A-1)  \label{e5}
\end{equation}

\begin{equation}
\sigma ^{\prime \prime }=-\left[ \frac{2}{x}+\frac{1}{2}\left( \frac{%
B^{\prime }}{B}-\frac{A^{\prime }}{A}\right) \right] \sigma ^{\prime
}-A\sigma \left( \frac{w^{2}}{B}-1\right)   \label{e6}
\end{equation}

where primes means $d/dx$ and:

\smallskip 
\[
x=mr\,\,,\,\,\sigma (x)=\phi (r)\sqrt{8\pi G}\,\,,\,\,w=\frac{\omega }{m}
\]

\subsection{Putting on Fermions}

According to Chandrasekhar \cite{Chandrasekhar39}, an equation of state can 
describe a perfect fluid of degenerate Fermi gas, with density and pressure 
given by:

\[
\rho =K(\sinh t-t)
\]

\[
p=\frac{K}{3}\left( \sinh t-8\sinh \frac{t}{2}-3t\right) 
\]

Where $K=m_{n}^{4}/32\pi ^{2}$, and $m_{n}$ is the fermion mass.

The parameter $t$ is related to the maximum momnetum in the Fermi
distribution.

Oppenheimer and Volkoff \cite{Oppenheimer39} found equilibrium configurations by using
Einstein equations with $T_{1}^{1}=T_{2}^{2}=T_{3}^{3}=p$ and $%
T_{0}^{0}=-\rho $, and the hydrostatic equilibrium equation for the pressure:
\[
p^{\prime }=-\frac{(\rho +p)B^{\prime }}{2}\frac{B^{\prime }}{B}
\]

Numerical evolution of these equations gives rise to a compact
configuration: given an initial value $t_{0}$ the function $t(r)$ decreases
and reaches zero for a finite value of $r$. 

Boson-Fermion Stars have been originally proposed by Henriques, Liddle \&
Moorhouse \cite{Henriques89}. 

The energy-momentum is composed with bosonic (B) and fermionic (F)
counterparts:
\[
T_{\mu \nu }=T_{\mu \nu }^{B}+T_{\mu \nu }^{F}
\]
where:

\[
T_{\mu \nu }^{B}=\partial _{\nu }\Phi ^{*}\partial _{\mu }\Phi +\partial
_{\mu }\Phi ^{*}\partial _{\nu }\Phi -g_{\mu \nu }\left( \partial _{\lambda
}\Phi ^{*}\partial ^{\lambda }\Phi +m^{2}\Phi ^{*}\Phi \right) 
\]
\[
T_{\mu \nu }^{B}=(\rho +p)u_{\mu }u_{\nu }+g_{\mu \nu }p
\]
And the set of equations for a boson-fermion star is:
\[
A^{\prime }=xA^{2}\left[ 2\overline{\rho }+\left( \frac{w^{2}}{B}+1\right)
\sigma ^{2}+\frac{\sigma ^{\prime \,2}}{A}\right] -\frac{A}{x}(A-1)
\]

\[
B^{\prime }=xAB\left[ 2\overline{p}+\left( \frac{w^{2}}{B}-1\right) \sigma
^{2}+\frac{\sigma ^{\prime \,2}}{A}\right] +\frac{B}{x}(A-1)
\]

\[
\sigma ^{\prime \prime }=-\left[ \frac{2}{x}+\frac{1}{2}\left( \frac{%
B^{\prime }}{B}-\frac{A^{\prime }}{A}\right) \right] \sigma ^{\prime
}-A\sigma \left( \frac{w^{2}}{B}-1\right) 
\]
\[
t^{\prime }=-2\frac{B^{\prime }}{B}\frac{\sinh t-2\sinh (t/2)}{\cosh
t-4\cosh (t/2)+3}
\]

where primes means $d/dx$ and:\smallskip 
\[
x=mr\,\,,\,\,\sigma (x)=\phi (r)\sqrt{8\pi G}\,\,,\,\,w=\frac{\omega }{m}
\]
\[
\overline{\rho }=\frac{4\pi G}{m^{2}}\rho (t)\,\,,\,\,\overline{p}=\frac{%
4\pi G}{m^{2}}p(t)
\]

\section{Rotating Boson-Fermion Stars}

\subsection{Rotating Fermion Stars}

Hartle \cite{Hartle67} analyzed slowly rotating relativistic stars of perfect fluid:
\[
ds^{2}=-e^{\nu }d\tau ^{2}+e^{\lambda }dr^{2}+r^{2}\left[ d\theta ^{2}+\sin
^{2}\theta \left( d\varphi -Ld\tau \right) ^{2}\right] 
\]
where $L$, $\nu $ and $\lambda $ are funcions of $r$ and $\theta $.
Considering $R$ the average radius and $\Omega $ the angular velocity as
seen by an observer at infinity, slow rotation means $R\Omega \ll c$.

We use expansions up to order $\Omega ^{2}$. In this case, one can
approximate energy and pressure by using:
\[
E=\rho +O(\Omega ^{2})
\]
\[
P=p+O(\Omega ^{2})
\]
since $L$ must be odd under the transformations, $\varphi \rightarrow
-\varphi $ and $\Omega \rightarrow -\Omega $, which preserves the metric
invariance under inversion of rotation.

In this approximation one can write:
\[
T_{\alpha \beta }^{\,\,F}=(E+P)u_{\alpha }u_{\beta }+g_{\alpha \beta
}P\simeq (\rho +p)u_{\alpha }u_{\beta }+g_{\alpha \beta }p
\]
and seeking solutions for $R_{3}^{0}=8\pi GT_{3}^{0}$ one can use $d\varphi
=\Omega d\tau $ to write $u^{3}=\Omega u^{0}$, and obtain:
\[
T_{3}^{0}=(E+P)(u^{0})^{2}(g_{03}+\Omega g_{33})
\]
\[
=(\rho +p)e^{-\nu }(\Omega -C)r^{2}\sin ^{2}\theta +O(\Omega ^{3})
\]

Calculations can be simplified by defining:
\[
\overline{C}=\Omega -C
\]
and using spherical harmonics expansion:
\[
\overline{C}(r,\theta )=\sum_{l=1}^{\infty }\overline{C}_{l}(r)\left[ -\frac{%
1}{\sin \theta }\partial _{\theta }P_{l}(\cos \theta )\right] 
\]

Adjustign the expansion coefficients to $\overline{C}(r,\theta )\simeq 
\overline{C}_{1}(r)$ and $\overline{C}$ becomes function of $r$ alone.

\subsection{Rotating Boson-Fermion Stars}

We \cite{Sousa01} analyzed slowly rotating relativistic stars composed with a
fermionic perfect fluid and a scalar filed. The equations, in this case up
to order $\Omega ^{2}$, read:

\[
A^{\prime }=xA^{2}\left[ 2\overline{\rho }+\left( \frac{w^{2}}{B}+1\right)
\sigma ^{2}+\frac{\sigma ^{\prime \,2}}{A}\right] -\frac{A}{x}(A-1)
\]

\[
B^{\prime }=xAB\left[ 2\overline{p}+\left( \frac{w^{2}}{B}-1\right) \sigma
^{2}+\frac{\sigma ^{\prime \,2}}{A}\right] +\frac{B}{x}(A-1)
\]

\[
\sigma ^{\prime \prime }=-\left[ \frac{2}{x}+\frac{1}{2}\left( \frac{%
B^{\prime }}{B}-\frac{A^{\prime }}{A}\right) \right] \sigma ^{\prime
}-A\sigma \left( \frac{w^{2}}{B}-1\right) 
\]
\[
t^{\prime }=-2\frac{B^{\prime }}{B}\frac{\sinh t-2\sinh (t/2)}{\cosh
t-4\cosh (t/2)+3}
\]
\[
C^{\prime \prime }=\left[ \frac{4}{x}-\frac{1}{2}\left( \frac{B^{\prime }}{B}%
-\frac{A^{\prime }}{A}\right) \right] C^{\prime }+4A(\overline{\rho }+%
\overline{p})\overline{C}
\]
Hence, up to order $\Omega ^{2}$ equations display no influence of parameter 
$C(x)$ over the fields $\sigma $, $t$, $A$ and $B$. Thus, for slow rotation,
if one finds an equilibrium configuration changes on the rotation parameter $%
\overline{C}$ cannot cause modifications on structural values (mass, radius
and number of particles). This means that one can obtain several slow
rotating configurations (several values of $\overline{C}$) for
boson-fermions stars that have the same structure ($\sigma $, $t$, $A$ and $B
$). Figure 1 shows the resulting plot for a known stable boson-fermion
configuration \cite{Sousa01} with $\overline{C}_{0}=\overline{C}_{1}(x=0)=1$%
, central density for bosons given by $\sigma _{0}=0.2$, and fermions
central parameter $t_{0}=4.0$. 

\begin{figure}
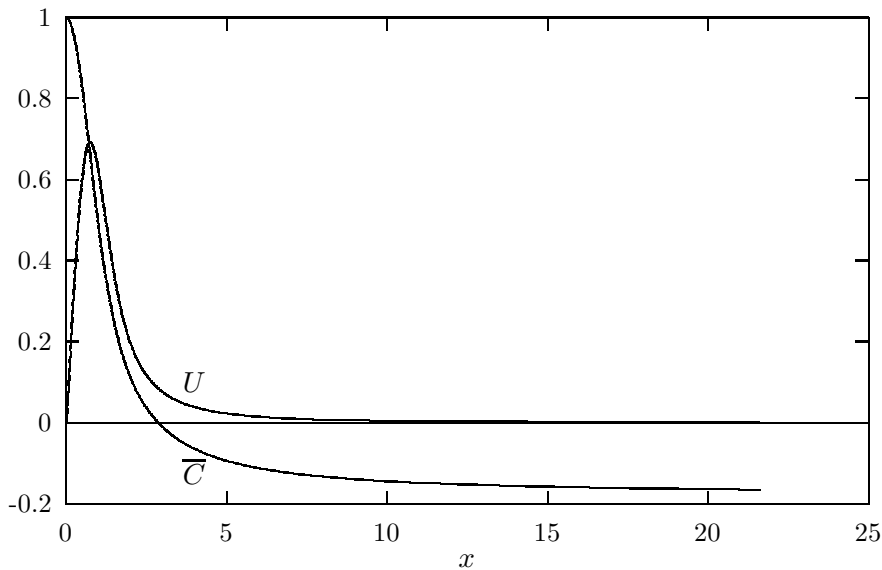

\setlength{\unitlength}{0.240900pt}
\ifx\plotpoint\undefined\newsavebox{\plotpoint}\fi

 \caption{Rotation parameter evolution for the case discussed in the text. 
  In this case $\lim_{x\rightarrow\infty}\overline{C}(x)=\Omega =-0.1651$.}
\end{figure}

\section{Discussion}
Using the Lagrangian for a scalar field coexisting with a fermionic fluid in a spherically symmetric configuration it has been possible to obtain stable rotating configurations for slowly rotating boson-fermion stars with the use of a perturbative prescription up to order $\Omega^2.$ Previously, Kobayashi {\em et al\/}~\cite{Kobayashi94} analyzed pure boson star slow rotation using relativistic prescription, and showed that perturbatively it is not possible to obtain rotational modes for boson stars. In contrast, after that \cite{Silveira95} it has been found boson star slow rotation by considering the quantum nature of the scalar field with axial symmetry. In this case it is possible to obtain stable configurations with $l\not= 0$ using Newtonian approach. The reason for the difference between our results and those obtained by Kobayashi {em et al\/} might be related to the quantization of the total angular momentum, as reported by Mielke and Schunk~\cite{Mielke96}.

Future projects could search for pure boson stars rotation using relativistic prescription. It is also of interest to determine the differences between the two {\em ansatz} studied up to now (with and without equatorial symmetry). Exploring expansions over $\Omega^2$ case could give informations concerning deformations on boson star's surface. It could be also of great interest to determine the gravitational radiation from decays between two rotational states.~\cite{Ryan97}

\end{document}